\documentclass{aastex}
\usepackage{emulateapj5}
\usepackage{float}

\slugcomment{Printed \today}

\shorttitle{The Complex Mid-Infrared Structure of IRAS 20126+4104}
\shortauthors{De Buizer}

\begin{document}

\title{The Complex Mid-Infrared Structure at the Heart of IRAS 20126+4104}

\author{James M. De Buizer}
\affil{Gemini Observatory, Casilla 603, La Serena, Chile;
jdebuizer@gemini.edu}

\begin{abstract}

The mid-infrared emission at the center of IRAS 21026+4104 is not
that of a simple compact source, as one would expect from an
isolated high mass protostellar object. Furthermore the central
thermal infrared emission does not appear to be coming directly from
a circumstellar disk as has been recently hypothesized from
near-infrared observations. The mid-infrared structure is complex,
but with the help of multiple wavelength information two plausible
scenarios to explain the emission in the region are advanced. The
first is that there is a tight cluster of young stellar objects
here. The second is that the mid-infrared emission and masers are
delineating the walls of the outflow cavities of a massive stellar
source located in the center of the near- and mid-infrared dark
lane.

\end{abstract}

\keywords{circumstellar matter --- infrared: ISM
--- ISM: individual (\objectname{IRAS 20126+4104}) --- stars: formation
--- stars: early type}

\section{Introduction}

It is not clear if massive stars in general form through accretion
processes similar to low-mass stars. At present, the best example of
a massive star with a possible accretion disk and bipolar outflow is
considered to be IRAS 20126+4104 (see Cesaroni et al. 2006 for
review). Originally thought to be a simple and compact high mass
protostellar object (a.k.a. hot molecular core), more recent
observations by Cesaroni et al. (2005) have shown this source to be
$\sim$7 M$_{\odot}$, surrounded by a $R\sim$5000 AU Keplerian
rotating CS ``disk'', with an outflow near perpendicular seen in
H$_2$ and SiO. Whether this large ``disk'' is a circumstellar
accretion disk or a accretion envelope feeding some unseen central
accretion disk is still not understood.

Recent observations in the near-infrared (NIR) by Sridharan et al.
(2005) revealed two K-band sources separated by a ``dark lane'',
interpreted as scattered emission on either side of a silhouette
disk at the center of the larger rotating molecular ``disk''. It was
also concluded that the thermal emission from the 5 $\micron$
observations may be tracing the dust emission from the central
accretion disk itself.

Observations at longer thermal infrared wavelengths are not hindered
as much by extinction so are better tracers of the thermal dust
emission that may be coming from the disk. In this article I
describe recent high angular resolution mid-infrared (MIR)
observations that reveal a much more complex dust structure than
previously thought.

\section{Observations}

Observations were performed at Gemini North with the Michelle MIR
imager and spectrograph. The instrument uses a Raytheon
320$\times$240 pixel Si:As IBC array with a pixel scale of
0.099$\arcsec$ pixel$^{-1}$. Sky and telescope subtraction were
achieved through the standard chop-nod technique. Imaging was
performed using the \emph{Si-6} ($\lambda$$_c$=12.5 $\micron$,
$\Delta\lambda$=1.2 $\micron$) filter and the \emph{Qa}
 ($\lambda$$_c$=18.3 $\micron$, $\Delta\lambda$=1.6
$\micron$) filter. On-source exposure times of 282s and 82s were
used, respectively. Flux calibration was achieved by observing the
MIR standard star HD199101 at a similar airmass to IRAS 20126+4104.
Assumed flux densities for HD199101 were taken to be 5.84 Jy at 12.5
$\micron$ and 2.63 Jy at 18.3 $\micron$. Flux densities measured are
listed in Table 1. Uncertainties in these flux densities are
estimated to be 10\% at 12.5 $\micron$ and 15\% at 18.3 $\micron$.

The images at both wavelengths were deconvolved using the maximum
likelihood method (Richarson 1972, Lucy 1974). Since no natural
point-spread function (PSF) stars were observed, artificially
generated PSFs with full width at half maximums equal to the
diffraction limit (0.33$\arcsec$ at 12.5$\micron$ and 0.48$\arcsec$
at 18.3$\micron$) were used in the deconvolutions. The deconvolution
routine was stopped at 25 iterations for the 12.5 $\micron$ image
and 50 iterations for the 18.3 $\micron$ image. These deconvolved
images compare favorably to simple unsharp masking of the original
images, and thus the revealed sub-structures in the deconvolved
images are believed to be real.

The relative astrometry between the NIR (K, L and M) and the MIR
images was accurately determined due to the presence of two (M) or
three (K and L) common point sources in the field.  The mean
residual offset between the 12.5 $\micron$ and NIR images was found
to be less than a Michelle pixel (0.099$\arcsec$) in all cases
(0.04$\arcsec$ for K, 0.03$\arcsec$ for L, and 0.09$\arcsec$ for M).

\section{Results}

Figure 1 shows a portion of the deconvolved 12.5 $\micron$ field
centered on the coordinates of the 1.3 cm radio continuum peak of
Hofner et al. (2006).  Near the origin is a bright MIR double source
(IRAS20126+4104:D06 1 and IRAS20126+4104:D06 2, from now on these
will be referred to as simply D06 1 and D06 2). These MIR sources,
like the double source seen at K by Sridharan et al. (2005), have a
``dark lane'' separating them. Most importantly, the dark lane
between the K-band sources, which is hypothesized by Sridharan et
al. (2005) to be the ``absorbtion lane'' of the NIR silhouette disk,
is exactly coincident with the dark lane between the two
mid-infrared sources.

Figure 2 shows that the comparisons between the MIR images and the L
and M images of Sridharan et al. (2005). The M peak is coincident
with the MIR emission from D06 2. Consequently this means that the M
peak is not located in the dark lane as suggested by Sridharan et
al. (2005). Since this NIR dark lane is also the location of minimum

\begin{table}[H]
\begin{center}
\scriptsize \caption{Observed and Derived Parameters}
\begin{tabular}{lcccc}
\hline
Source & F$_{12.5\micron}$ &F$_{18.3\micron}$ & L$_{MIR}$$^a$ &Spec     \\
       &(Jy)               &(Jy)              &L($_{\odot}$)  &Type$^a$ \\
\hline
D06 1                 &0.71  &3.71  &58   &B9  \\
D06 2                 &0.34  &2.97  &63   &B9  \\
Double Source$^b$     &1.06  &6.45  &106  &B8  \\
Whole Field$^c$       &1.89  &23.5  &715  &B5  \\
\hline
\end{tabular}
\medskip \\
\end{center}
\scriptsize
 {\em Note:} Accuracy of the flux densities: 10\% at 12.5$\micron$, 15\% at 18.3 $\micron$.\\
 {\em $^a$} These are lower limits to the luminosity and spectral type
 derived from the MIR flux densities and assuming all emission is
 from a single stellar source. See De Buizer et al. (2005) for
 techniques and assumptions. \\
 {\em $^b$} Flux density in a 1$\farcs$0 radius aperture
 around D06 1 and D06 2 only.\\
 {\em $^c$} Flux density in a 8$\farcs$0 radius aperture
 encompassing all MIR sources.\\

\end{table}

flux density between the MIR sources D06 1 and D06 2, and D06 2 is
coincident with the M peak, the suggestion by Sridharan et al.
(2005) that the longer wavelength NIR emission is tracing thermal
dust emission from a near edge-on circumstellar disk inside the dark
lane between the K band sources is not confirmed.

The secondary peak seen at L by Sridharan et al. (2005) is found to
be coincident with a MIR extension to the south of the MIR double
source in the deconvolved 12.5 $\micron$ image (labeled the
``southern extension'' in Figure 1). Another MIR extension to the
west of the D06 1 is weak at 12.5 $\micron$, but is clearly evident
and resolved from the double source at 18.3 $\micron$ (labeled the
``western extension'' in Figure 1). One further difference between
the 12.5 and 18.3 $\micron$ deconvolved images is the shape of D06
1. At 18.3 $\micron$ the source is elongated further to the south
($\sim$0.2$\arcsec$) than at 12.5 $\micron$ (see Figure 2).

Figure 3 shows the comparison of the 18.3 $\micron$ deconvolved
image to several other wavelengths known to have very good
($\lesssim$0.15$\arcsec$) absolute astrometric accuracy. The K band
image (blue contours) of Sridharan et al. (2005) is claimed to be
known to an absolute astrometric accuracy of 0.15$\arcsec$. Because
the relative registration between the NIR and MIR images is shown to
be better than 0.1$\arcsec$, the absolute position error of the MIR
image is therefore the quadrature addition of these two values,
namely $<$0.18$\arcsec$.

Also shown in Figure 3 are the locations of OH, water, and methanol
masers from Edris et al. (2005). The absolute astrometry of these
masers are known to better than 25 mas. Also plotted in Figure 3 are
the three water maser locations of Tofani et al. (1995), which
appear to be associated with the locations of D06 1 and 2 and the
MIR western extension. The 3.6 cm radio continuum (white contours)
from Hofner et al. (2006) are also shown in Figure 3. The brightest
radio component has a peak near the southern elongation of D06 1
seen at 18.3 $\micron$. The westernmost radio continuum component
lies very close ($<$0.1$\arcsec$) to the MIR western extension. The
radio continuum component to the south has no associated MIR
emission.

Figure 3 shows a complex distribution of multi-wavelength emission,
however the astrometric precision of all of these observations
facilitates the interpretation of the combined data set. Two
possible scenarios will now be forwarded that may explain the
observed complexity, though further data will be require to test
either scenario. The first scenario is that there is a cluster of
young sources resent at the center of IRAS 210126+4104, and the second

\begin{figure}[H]

\epsscale{0.80} \plotone{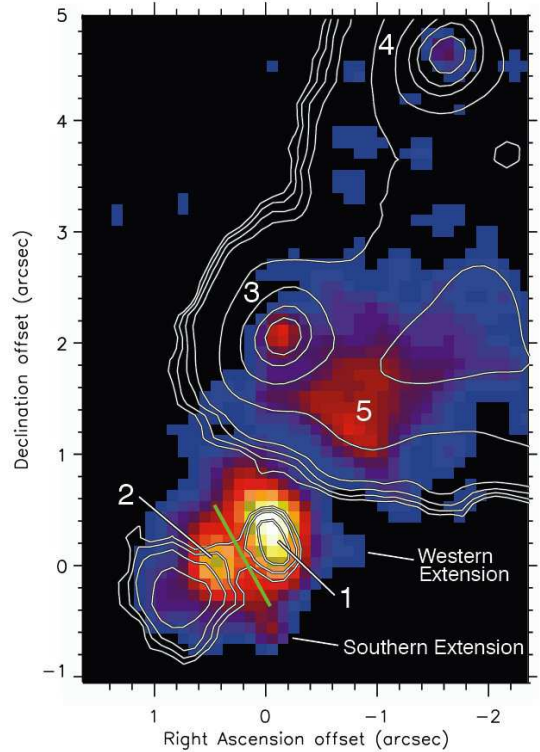}

\caption{The 12.5 $\micron$ deconvolved Michelle image of IRAS
20126+4104 in false color overlaid with the K band contours of
Sridharan et al. (2005). Three common sources (only two sources,
D06 3 and 4, are shown here) in the field were used to accurately
register the MIR and NIR images. The green line shows the location
of the NIR/MIR dark lane. The origin of the figure is the location
of the Hofner et al. (2006) 1.3 cm radio continuum peak (R.A.(J2000)
= 20$^h$ 14$^m$ $26\rlap.^s031$, Decl.(J2000) = $41^\circ$ $13'$
$32\farcs58$).} \label{fig1}

\end{figure}

is that the different wavelengths of emission are coming from
different tracers of the bipolar outflow from this region centered
on the dark lane seen in the NIR and MIR.

\subsection{Multiple Source Scenario}

The heart of IRAS 210126+4104 may contain a small cluster of young
stellar sources of various masses and in various stages of
formation. There could be several individual sources here, among
them could be D06 1, D06 2, each of the 3 radio continuum peaks.
These sources would have various associations, for instance, D06 2,
which is coincident with the peak seen by Sridharan et al. (2005) at
M, as seen in Figure 2. This source is also traced by copious
amounts of methanol, OH, and water maser emission, but has no radio
continuum emission of its own. Based solely on the MIR luminosity
and assuming emission from a single stellar source, a lower limit to
the spectral type for D06 2 of B9 was derived (Table 1) using the
methods of De Buizer et al. (2005). In contrast, the source
associated with the brightest radio continuum peak does not appear
to have any directly related MIR emission at 12.5 $\micron$, and
based on the work of Hofner et al. (1999) could be an embedded UCHII
region of a star of spectral type B3. However, the MIR emission of
D06 1 does extend towards this source at 18.3 $\micron$ and may be
marking the weak emission at longer MIR wavelengths from the radio
continuum source.  In this multiple source scenario, the dark lane
seen in the NIR and MIR would simply be a

\begin{figure}[H]
\epsscale{0.7} \plotone{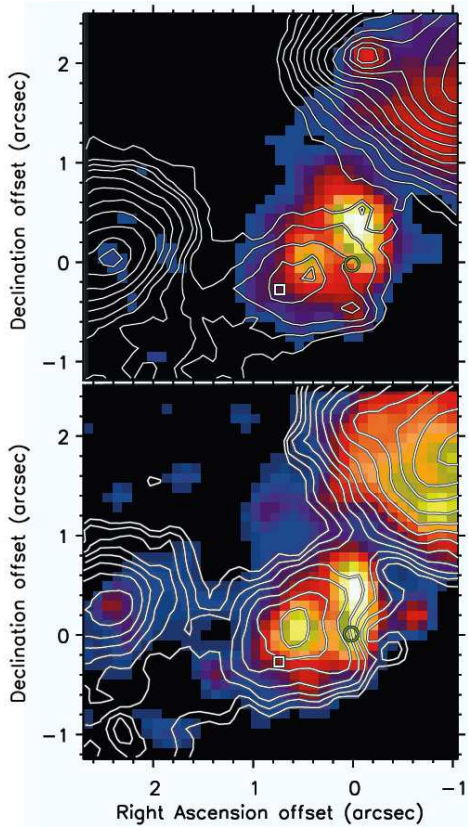}

\caption{Comparisons between MIR and NIR morphologies. Upper panel
shows the deconvolved 12.5 $\micron$ image in false color overlaid
with the L-band contours of Sridharan et al. (2005). The lower panel
shows the the deconvolved 18.3 $\micron$ image in false color
overlaid with the M-band contours of Sridharan et al. (2005). The
origins are the same as Figure 1. The locations of the 1.3 cm radio continuum peak (green circle) and the southeastern K band peak (white square) are shown.} \label{fig2}
\end{figure}

space separating the
various IR sources present, like the ``dark lane'' between D06 1 and
D06 5.

This scenario of having more than one young stellar source at the
heart of IRAS 20126+4104 could help to explain the detection of
apparently two outflows from this region: one nearly N-S and the
other NW-SE, and with red-shifted and blue-shifted lobes reversed in
the two cases (Cesaroni et al. 2005). These outflows could come from
two independent stellar sources, as has been suggested before (e.g.
Lebr\'{o}n et al. 2006). Given its central location with respect to
the overall NIR/MIR emission as well as the H$_2$ outflow, and its
association with the majority of the masers in the region, D06 2
could be the dominant young stellar object in the cluster in this
scenario.

Hofner et al. (1999) point out that if the three radio peaks are
considered to be three ZAMS stars of spectral type B3 (as indicated
by their radio continuum emission), there is a problem accounting
for the total luminosity of the region at far-infrared (FIR)
wavelengths. Three such stars combined would have a luminosity of
only 3.0$\times$10$^3$ L$_{\odot}$, but the FIR luminosity measured
by IRAS is 1.3$\times$10$^4$ L$_{\odot}$. But if just one of these
MIR sources without radio continuum emission (i.e. D06 2) is an
HMPO, then this could explain the higher flux densities at FIR
wavelengths since it is at these longer wavelengths that the SEDs of
HMPOs peak.

\subsection{IR Emission from Outflow Scenario}

Sridharan et al. (2005) assert that the dark lane between the K band
sources is an ``absorption lane'' of a silhouette disk of a young massive stellar source and that the
longer wavelength IR emission comes from the dust emission of the
accretion disk in this absorption lane. However, the NIR and MIR
emission share a common dark lane between the northwestern and
southeastern sources of IR emission. In this section an alternate
scenario is proposed in which this dark lane is indeed an absorption
lane of an embedded stellar source, however it is instead argued
that both the NIR and MIR are coming from scattered and direct
emission, respectively, off the outflow cavities from this stellar
source.

Figure 3 shows that all of the maser species are predominantly
situated along the inner edges of the absorption lane but are
associated with the infrared emission and not the lane itself
(contrary to what is claimed by Edris et al. 2005 based on the NIR
data alone). Therefore a more simplistic scenario could be that the
masers and MIR emission are delineating the walls of the cavities of
an outflow coming from the massive embedded star at the center of
the absorption lane. The NIR, MIR, radio continuum, and maser
emission in this region would be coming from different parts of the
NW and SE outflow cavities. The outflow axis seen in H$_2$, SiO, and
HCO$^+$ (Cesaroni et al. 1999) has a position angle close to
$\theta$$\sim$-60$^{\circ}$. The dark lane seen in the MIR lies at
$\theta$$\sim$30$^{\circ}$ (Figure 1), perpendicular to the outflow.
The other extended MIR emission components in the field
(IRAS21026+4104:D06 5, and perhaps D06 3 and D06 4 in Figure 1) lie
close to this outflow axis and may be outflow-related MIR
emission since they are coincident with K and H$_2$ emission here.

The two northern knots of radio continuum emission in Figure 3 are
hypothesized by Hofner et al. (2006) to perhaps be a part of a
outflow jet emanating from a stellar source located in the IR dark
lane. This jet is aligned slightly more E-W
($\theta$$\sim$-70$^{\circ}$) than the outflow seen in H$_2$, SiO,
and HCO$^+$, and is much different than the larger scale CO outflow
(Lebr\'{o}n et al. 2006) that appears closer to N-S
($\theta$$\sim$-25$^{\circ}$). The high shock temperature required to excite the H$_2$ $\upsilon$=1-0 S(1) transition is several thousand degrees, whereas for CO (J=2-1) it is just a few K. H$_2$ rapidly cools below the temperature where it can emit in a time scale of $\sim$1 yr (Le
Bourlot et al. 1999). However, the CO remains excited in the post-shock gas (since the gas temperature here is in excess of the $\sim$5K needed for this to happen), thus providing a fossil record of the outflow shock-interaction. Therefore instead of being due to multiple outflowing sources as described in the last section, there may be
one wandering outflow responsible for the apparent array of outflow
angles seen. The CO could be indicating the past history of the outflow,
whereas the H$_2$ and radio continuum emission is denoting what is
occurring more recently.

In this outflow cavity scenario, the southeastern component of the
double source at K and the MIR emission from D06 2 would be tracing
different parts of the southeastern outflow cavity. Cesaroni et al.
(2005) finds that the high-velocity outflowing gas is red-shifted to
the southeast and blue-shifted to the northwest. Therefore the disk
is not exactly edge-on, and we expect to have a more direct

\begin{figure}[H]
\epsscale{0.85} \plotone{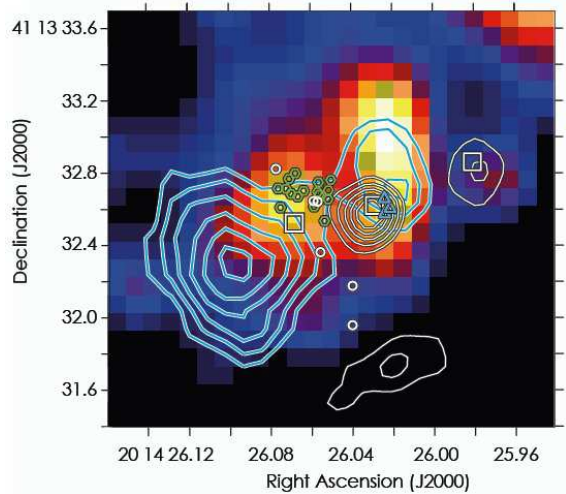}

\caption{The deconvolved 18.3 $\micron$ image in false color is
overlaid with contours from several wavelengths. The blue contours
are the K-band contours of Sridharan et al. (2005), the white
contours are the 3.6 cm radio continuum emission from Hofner et al.
(2006, in prep). The water masers (white squares) of Tofani et al.
(1995) are shown, as well as the OH (white circles), water (blue
triangles) and methanol masers (green pentagons) from Edris et al.
(2005).} \label{fig3}
\end{figure}

and
unextinguished view of the northwestern outflow cavity than the
southeastern. The southeastern NIR and MIR emission is extended on
the same scale, however the MIR peak is closer to the absorption
lane than the NIR peaks. The emission in the MIR of D06 2 may be seen
closer to the dark lane because the MIR emission penetrates more
easily through obscured areas and is generally orders of magnitude less
efficiently scattered than K emission in an outflow cavity. This holds true under the general assumption that there is not a large variation in the dust composition and size distribution throughout this whole region. Assuming this to be true, the emission in the thermal IR (MIR and M) in the southeastern cavity
would appear closest to the absorption lane, at L the peak is further
out, and at K even further out.

Given the more favorable viewing angle and with extinction playing
less of a role, we have a more direct view of the northwestern
outflow cavity. This means we can directly see the hot dust down
closer to the stellar source itself at L and M. This, plus
resolution effects due to the fact that the emission from above and
below the dark lane are closer together at L and M than at K, may
account for the fact that the extinction lane is not as prominent at
these wavelengths. The K emission may be seen further out due to the
efficiency with which it scatters, and MIR emission may be further out
because it is tracing the cooler dust. This, combined with the
better resolution at these wavelengths may help to enhance the
appearance of the dark lane at K and in the MIR.

The MIR southern extension may also be better explained by the outflow
cavity scenario. The emission at this location is not apparent at K,
bright at L, marginally detected at M, present at 12.5 $\micron$,
and shifted in location at 18.3 $\micron$. If the emission is coming
from a outflow region, it is plausible that line emission, temperature
and optical depth gradients, and scattering could cause the infrared
emission peaks to move as a function of wavelength within the
outflow cavity. However further data, including spectroscopy, will be required to know the true nature of the southern extension. The main point however is that this behavior as a function of wavelength could not
be due to a stellar ``binary companion'', as suggested by Sridharan et al. (2005),
as dusty young stellar sources have increasing flux density as a
function of increasing wavelength.

The outflow scenario also fits well the observed maser properties.
Moscadelli et al. (2000, 2005) have already shown that their
observations of the water maser spots can be well-fit to a model
where they are moving outward from the location of the dark lane
along the axis of outflow. In a similar fashion, here it is
suggested that the methanol maser emission is associated with the
inner parts of the southeastern outflow cavity. This matches recent
observations indicating that methanol masers are associated with the
outflows from young massive stars (e.g., De Buizer 2003, De Buizer
2006).

Assuming the unseen stellar source at the center of the dark lane is
responsible for radio continuum emission in outflow, Hofner et al.
(1999) derived a B1 spectral type for this central star under the
assumption that the radio continuum emission is beamed emission into
the jet cavity.  All of the extended MIR emission could indeed be heated by
a single B1 star as well. The farthest MIR source, D06 4, is $\sim$8500 AU away from the radio
continuum peak, but assuming smooth astronomical silicates and small
grains, it is possible to heat dust to $\sim$120K out to this
distance (see De Buizer 2006). Therefore, all of the MIR emission
observed to be associated with IRAS 21026+4104 requires no more than
a single B1 stellar source at its heart, and a multiple source scenario is not necessary. However, it is possible that there are other young stellar sources (perhaps D06 3, 4, and 5) present and helping to heat the dust in the regions farther from the center.

\acknowledgments Based on observations obtained at the Gemini
Observatory, which is operated by the AURA, Inc., under a
cooperative agreement with the NSF on behalf of the Gemini
partnership: NSF (US), PPARC (UK), NRC (Canada), CONICYT (Chile),
ARC (Australia), CNPq (Brazil) and CONICET (Argentina). Gemini
program identification number associated with the data is
GN-2005B-Q-21.

\end{document}